\documentclass[letterpaper,pra,aps,twocolumn,showpacs,superscriptaddress]{revtex4}
\usepackage{bm,color}
\usepackage{mathptmx}
\usepackage[T1]{fontenc}
\usepackage[utf8]{inputenc}
\usepackage{graphicx} 
\usepackage{amsmath} 
\usepackage{amsfonts} 
\usepackage[caption=false]{subfig} 

\newcommand{\bea}[1]{\begin{align}#1\end{align}}

\newcommand{\com}{\,\text{,}}
\newcommand{\dt}{\,\text{.}}

\newcommand{\bi}{\begin{itemize}}
\newcommand{\ei}{\end{itemize}}

\newcommand{\bbm}{\begin{pmatrix}}
\newcommand{\ebm}{\end{pmatrix}}
\newcommand{\bma}{\begin{matrix}}
\newcommand{\ema}{\end{matrix}}
\newcommand{\bsm}{\begin{smallmatrix}}
\newcommand{\esm}{\end{smallmatrix}}
\newcommand{\bsbm}{\left( \begin{smallmatrix}}
\newcommand{\esbm}{\end{smallmatrix}\right)}

\newcommand{\up}{\uparrow}
\newcommand{\down}{\downarrow}
\newcommand{\vb}[1]{\left( #1 \right)}
\newcommand{\vsb}[1]{\left[ #1 \right]}
\newcommand{\vcb}[1]{\left\{ #1 \right\}}
\newcommand{\ave}[1]{\left\langle #1 \right\rangle}
\newcommand{\nn}[1]{\ave{#1}_{\rm n}}
\newcommand{\nnn}[1]{\ave{#1}_{\rm nn}}

\newcommand{\mbf}[1]{\mathbf{#1}}
\newcommand{\abs}[1]{\vert #1 \vert}
\newcommand{\abss}[1]{\vert #1 \vert^2}

\newlength{\paragraphskipping}
\newcommand{\eqb}{\nonumber\\ &\quad}

\newlength{\figurewidtha}
\newlength{\figureheight}
\begin{document}

\title{Sudden-quench dynamics of Bardeen-Cooper-Schrieffer states in deep optical lattices}
\author{Marlon Nuske}
\affiliation{Zentrum f\"ur Optische Quantentechnologien and Institut f\"ur Laserphysik, Universit\"at Hamburg, 22761 Hamburg, Germany}
\affiliation{National Institute of Standards and Technology, Gaithersburg, Maryland 20899, USA}
\author{L. Mathey}
\affiliation{Zentrum f\"ur Optische Quantentechnologien and Institut f\"ur Laserphysik, Universit\"at Hamburg, 22761 Hamburg, Germany}
\affiliation{The Hamburg Centre for Ultrafast Imaging, Luruper Chaussee 149, Hamburg 22761, Germany}
\author{Eite Tiesinga}
\affiliation{Joint Quantum Institute and Center for Quantum Information and Computer Science, National Institute of Standards and Technology and University of Maryland, Gaithersburg, Maryland 20899, USA}

\begin{abstract} 
We determine the exact dynamics of an initial Bardeen-Cooper-Schrieffer (BCS) state of ultra-cold atoms in a deep hexagonal optical 
lattice. The dynamical evolution is triggered by a quench of the lattice potential, such that the interaction strength $U_f$ is much larger than the hopping amplitude $J_f$. The quench initiates collective oscillations with frequency $\abs{U_f}/(2\pi)$ in the momentum occupation numbers and imprints an oscillating phase with the same frequency on the BCS order parameter $\Delta$. 
The oscillation frequency of $\Delta$ is not reproduced by treating the time evolution in mean-field theory. In our theory, the momentum noise (i.e. density-density) correlation functions oscillate at frequency $\abs{U_f}/2\pi$ as well as at its second harmonic.
For a very deep lattice, with zero tunneling energy, the oscillations of momentum occupation numbers are undamped. 
Non-zero tunneling after the quench leads to dephasing of the different momentum modes and a subsequent damping of the oscillations. The damping occurs even for a finite-temperature initial BCS state, but not for a non-interacting Fermi gas.
Furthermore, damping is stronger for larger order parameter and may therefore be used as a signature of the BCS state.
Finally, our theory shows that the noise correlation functions in a honeycomb lattice will develop strong anti-correlations near the Dirac point.
\end{abstract}

\date{\today}


\maketitle


\section{Introduction}
	Ultracold atoms in optical lattices are a versatile tool to simulate solid state phenomena \cite{bloch_many-body_2008}. 
	The tunability of lattice properties over a wide range of parameters is not only allowing experiments to explore regions of the phase diagram not attainable in solid state systems, but it also offers new, highly controllable methods for initiating dynamics. This has been extensively used for studying non-equilibrium dynamics in bosonic systems \cite{dziarmaga_dynamics_2010,morsch_dynamics_2006,polkovnikov_textitcolloquium_2011}. In particular, quenches of the lattice depth have been used to study the collapse and revival of a Bose-Einstein condensate (BEC) \cite{greiner_collapse_2002,mahmud_bloch_2014}. 

	At low temperatures fermionic atoms in optical lattices undergo a phase transition to a BEC of molecules for repulsive interactions and the paired Bardeen-Cooper-Schriffer (BCS) state \cite{bardeen_theory_1957,leggett_quantum_2006,koponen_sound_2006} for small attractive interactions \cite{sa_de_melo_crossover_1993,ketterle_making_2008}. 
	In the BCS regime the density and momentum distribution is nearly independent of the size of the order parameter (gap). It has therefore been proposed by Altman et al. to use the density-density correlation to measure the order parameter in experiment \cite{altman_probing_2004}. Greiner et al. have demonstrated that measuring the shot noise in absorption images makes the density-density correlations experimentally accessible \cite{greiner_probing_2005}. This has motivated several further studies of the density-density correlations \cite{lamacraft_particle_2006,belzig_density_2007,kudla_pairing_2015} as well as proposals to use them in order to distinguish different phases of ultracold fermions \cite{carusotto_coherence_2004,paananen_noise_2008,mathey_noise_2008,mathey_noise_2009}.
	An alternative approach to measuring the correlations in a Fermionic gas is to observe the time-evoulution after a quench of either the lattice depth or the interactions between atoms. In fact, Volkov and Kogan have predicted oscillations of the order parameter (gap) in the BCS regime over 40 years ago \cite{volkov_collisionless_1973}. Recently, this topic has attracted new attention and several different quenches of the interaction strength from a non-interacting state to the BCS regime \cite{barankov_collective_2004,tomadin_nonequilibrium_2008}, within the BCS regime \cite{warner_quench_2005,yuzbashyan_nonequilibrium_2005,yuzbashyan_solution_2005,barankov_synchronization_2006,dzero_spectroscopic_2007} and between the BCS and the BEC regime \cite{andreev_nonequilibrium_2004,szymanska_dynamics_2005,yuzbashyan_dynamical_2006,yuzbashyan_relaxation_2006,bulgac_large_2009,scott_rapid_2012} have been analyzed. 
	Phase diagrams of the asymptotic behaviour for long times after the quench have been obtained in  \cite{yuzbashyan_quantum_2015}. All of these theoretical models for quenches in Fermionic systems have in common that they use mean-field theory for both the initial state as well as the time evolution.

	The experimental realization of loading ultracold bosons \cite{soltan-panahi_multi-component_2011} and fermions \cite{tarruell_creating_2012,uehlinger_artificial_2013} into topological lattices, here the honeycomb (graphene) lattice, in particular, has started much interest in the exotic phase diagrams of these systems \cite{zhao_bcs-bec_2006,lee_attractive_2009,gremaud_pairing_2012,jotzu_experimental_2014}. 
	Furthermore, it was demonstrated that initiating dynamics in topological lattices gives direct experimental access to the band structure \cite{uehlinger_double_2013} as well as topological quantities such as chern numbers \cite{aidelsburger_measuring_2015}, the Berry curvature \cite{flaschner_experimental_2015} and Wilson lines \cite{li_experimental_2015}.

	In this paper we investigate the time evolution of a BCS state in the honeycomb lattice after a sudden ramp of the lattice depth. 
	We consider the Fermi-Hubbard model away from half filling for small attractive interactions. The corresponding ground state 
	is well described by mean-field BCS theory \cite{zhao_bcs-bec_2006}. 
	By exploiting the integrability of the BCS model we compute the full time evolution beyond mean-field theory for ramps to large final lattice depths, where the dynamics is determined by the interaction strength $U_f$ between the atoms, while the hopping strength $J_f$ is negligible. The quench is considered sudden with respect to many particle physics, but slow compared to the time scales of inter-band transitions. This regime is indeed achievable as we find that transitions between the lowest two bands are highly suppressed for a ramp of the lattice depth. Transitions to higher bands are negligible due to the large energy gap between bands.

	We find collective sinusoidal oscillations of the momentum occupation numbers with the frequency $\abs{U_f}/(2\pi)$ for all momentum modes. We also find that the phase of the complex-valued order parameter $\Delta(t)$ increases linearly in time, while its amplitude is time independent. In a Fermi-Hubbard model a quench of the lattice depth is formally equivalent to a quench of the interaction  strength. References \cite{yuzbashyan_relaxation_2006,barankov_synchronization_2006,scott_rapid_2012} studied such an interaction quench within the framework of Bogoliubov-de Gennes mean-field theory and predict that the time evolution of $\Delta(t)$ has large-amplitude, non-trivial oscillations. Such difference in predictions for the time dependence should be experimentally verifiable.

	We extend our analysis to include a small, finite tunneling energy after the quench. This leads to dephasing between different momentum modes and a subsequent damping of the oscillations. For times much smaller than $1/(\abs{U_f}^2J_f)^{1/3}$ we find a regime where damping occurs for an interacting initial state with a finite order parameter $\Delta$, while a non-interacting initial state does not damp. This motivates the use of the damping signal as an experimentally-accessible signature of $\Delta$. Fully numerical calculations with small systems using exact diagonalization show, however, that the oscillations may also damp for an interacting initial state with zero order parameter. In an experiment it may therefore be challenging to isolate the damping origin. 

	As a direct measure of pair correlations we also investigate the time evolution of the density-density correlations.
	For the BCS ground state these correlations are non-zero only for opposite momenta and can be used to estimate the size of the order parameter. Mean-field theory enforces that even after the ramp the correlations are only non-zero for those momenta while our exact theory predicts small corrections to these results.
	The discrepancy between mean-field and the exact theory becomes particularly strong at the Dirac points of the honeycomb lattice, where the first and second band touch linearly.

	The remainder of the article is set up as follows. Section \ref{sec:hamiltonian} describes the Hamiltonian as well as the initial state used for our calculations. We give the model for the time-evolution procedure in Sec.~\ref{model} and present the results in Sec.~\ref{oscillations}. In particular, Secs.~\ref{mom0} and \ref{momfin} describe the time evolution of the momentum modes for zero and finite hopping after the ramp, respectively, and Sec.~\ref{gap} describes the time evolution of the order parameter for both cases. The time evolution of higher-order correlation functions is analyzed in Sec.~\ref{highOrder}. Finally, we summarize in Sec.~\ref{sec:summary}.

\section{Hamiltonian and BCS ground-state}\label{sec:hamiltonian}

\begin{figure}[ht]
	   \centering
	   \includegraphics[width=\linewidth]{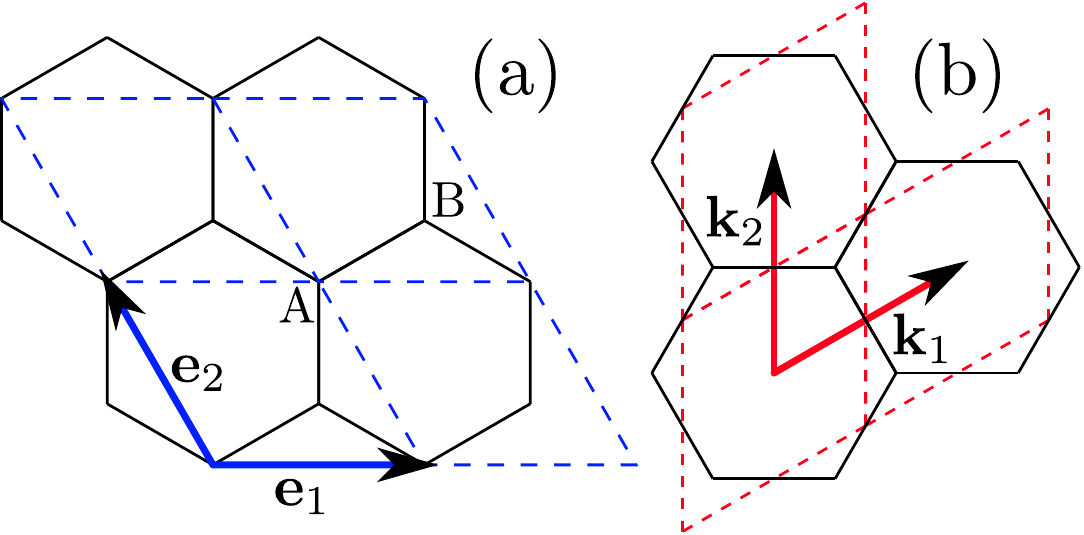}
	  \includegraphics[width=0.34\linewidth]{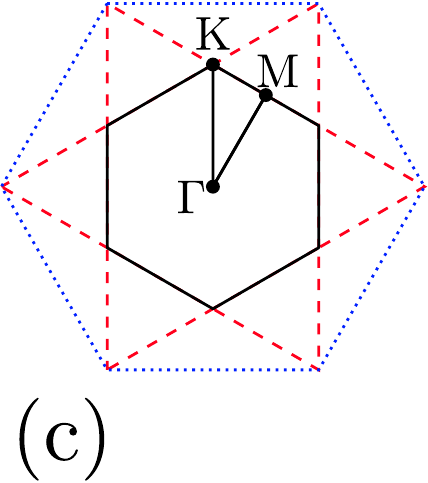}
	  \includegraphics[width=0.65\linewidth,trim=4.5cm 17.3cm 3.3cm 4.7cm]{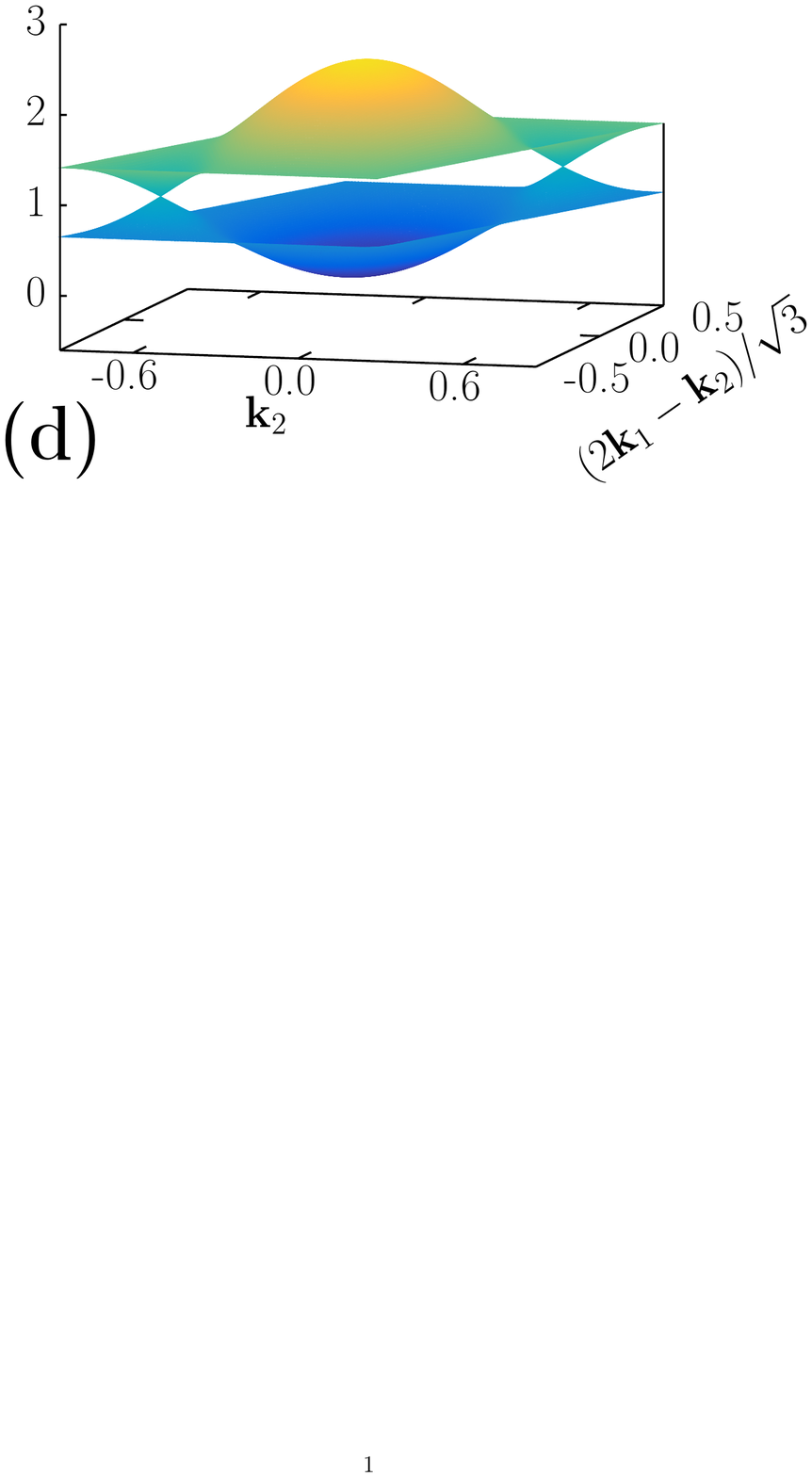}%
	   \caption{(color online) (a) Sketch of the honeycomb lattice with unit vectors $\mbf e_1=(a,0)^T$ and $\mbf e_2$ and lattice constant $a$. Solid black lines show the hexagonal lattice of Wigner-Seitz unit cells. The dashed blue parallelograms show an equivalent lattice spanned by $\mbf e_1$ and $\mbf e_2$. Both lattices contain two sites, $A$ and $B$, per unit cell. (b) Reciprocal lattice of the honeycomb lattice. Solid black hexagons show the first Brillouin zones at each lattice point and dashed red lines show the lattice spanned by the two reciprocal lattice vectors $\mbf k_1$ and $\mbf k_2$. (c) First, second and third Brillouin zones of the honeycomb lattice. Their boundaries are indicated by solid black, dashed red and dotted blue lines, respectively. The $\Gamma$, $M$ and Dirac ($K$-) points are marked by black dots. (d) Single-particle band structure of the honeycomb lattice for the first two bands and all quasi-momenta inside the unit cell spanned by the reciprocal lattice vectors $\mbf k_1$ and $\mbf k_2$ as defined in (b). 
	   The bands touch at the two Dirac points, which feature a linear dispersion relation.}
	   \label{fig:lattice}
	\end{figure}

	For our calculations we use a two-band attractive Fermi-Hubbard model with equal spin populations on a honeycomb lattice with on-site interactions, nearest and next-nearest neighbour hopping. Brillouin zones and lattice vectors in coordinate and reciprocal lattice space are defined in Fig.~\ref{fig:lattice}. The Hamiltonian in momentum space is given by
	\bea{
	H(\mu,J,J',U)&=H_J+H_U+H_\mu\label{eq:Hamiltonian}
	}
	with
	\bea{
	H_J&=\sum_{k} \vsb{ \epsilon_k \vb{a_{k,A}^\dagger a_{k,B}+b_{k,A}^\dagger b_{k,B}}+c.c.}\eqb
	+\sum_{k,C} \epsilon'_k \vb{a_{k,C}^\dag a_{k,C}+b_{k,C}^\dag b_{k,C}}\\
	H_U&=U\frac 1M \sum_{kpq,C}a_{k,C}^\dag a_{k+p-q,C} b_{p,C}^\dag b_{q,C}\\
	\text{and}\quad H_\mu&=-\mu \sum_{k,C}  \vb{a_{k,C}^\dag a_{k,C}+b_{k,C}^\dag b_{k,C}}\label{eq:HamMomChemPot}
	}
	where $H_J$ denotes the hopping part, $H_U$ the interaction part and $H_\mu$ the chemical potential part of the Hamiltonian. The operators $a_{kC}^\dag$ ($b_{kC}^\dag$) and $a_{kC}$ ($b_{kC}$) create and annihilate a spin down (up) fermion with quasi-momentum $\mbf k$ on the sublattice $C=A,B$. Here and throughout this paper, quasi-momentum sums run over all $M$ momenta in the first Brillouin zone, $M$ is the total number of lattice sites and $A$ and $B$ denote the two distinct lattice sites per unit cell as defined in Fig.~\ref{fig:lattice}(a). The interaction strength is given by $U<0$, while $J>0$ and $J'>0$ denote the hopping strengths between nearest and next-nearest neighbours, respectively. For a spin-balanced gas the chemical potential $\mu$ of the two species is equal. Finally,
	\bea{
	\epsilon_{k}(J) &= -J\vb{1+e^{-i\mbf k \cdot \mbf e_1}+e^{-i\mbf k \cdot \mbf e_3}} \equiv \abs{\epsilon_k} e^{i\phi_k}\\
	\epsilon'_k(J')&=-2J'\Big[\cos\vb{\mbf k\cdot \mbf e_1}+ \cos\vb{\mbf k\cdot  \mbf e_2 }+\cos\vb{\mbf k\cdot  \mbf e_3 }\Big]\com
	}
	where $\mbf e_3=\mbf e_1+\mbf e_2$. All operators obey fermionic anticommutation relations, e.g. $\{a_{p,C},a_{k,D}^\dagger\}=\delta_{pk}\delta_{CD}$. While the Fermi-Hubbard model is based on the lattice space Hamiltonian, given in App.~\ref{timeEvolCalc}, the above momentum space one is obtained by using the Fourier transforms from Eq.~\ref{fourier}.

	\graphicspath{{figures/}}

	The non-interacting Hamiltonian $H_J+H_\mu$ is exactly solvable by a unitary transformation to the operators $a_{k,1}^\dag$ ($a_{k,2}^\dag$) and $b_{k,1}^\dag$ ($b_{k,2}^\dag$) creating a fermion in the first (second) band. Their eigenenergies are spin independent and given by $\epsilon_{k,1}=\epsilon_k'-\mu-\abs{\epsilon_k}$ and $\epsilon_{k,2}=\epsilon_k'-\mu+\abs{\epsilon_k}$. The band structure for these two bands is shown in Fig.~\ref{fig:lattice}(d). 
	There has been much interest in the two Dirac \mbox{($K$-)} points, marked in Fig.~\ref{fig:lattice}(c), where $\epsilon_k=0$ and the two bands touch. The $K$-points feature a linear dispersion relation \cite{tarruell_creating_2012,lee_attractive_2009} and $\phi_k$, the complex phase of $\epsilon_k$, jumps by $\pi$ when going through the Dirac points in an arbitrary direction. The discontinuity in $\phi_k$ results in a non-zero Berry curvature or equivalently a non-zero Berry phase for any closed loop containing one of the Dirac points \cite{zhang_experimental_2005}.

	The low energy spectrum of $H$ is to good approximation given by that of the mean-field Hamiltonian
	\bea{
	H_{\text{mf}}(J,J',U,\mu)&=H_J+H_\mu +\sum_{k,C} \vb{\Delta^* a_{k,C}b_{-k,C} +\Delta b^\dag_{-k,C} a^\dag_{k,C}}\dt \label{meanFieldH}
	}
	with order parameter or gap $\Delta$, based on pairing between fermions of opposite spin and momentum \cite{bardeen_theory_1957,leggett_quantum_2006,koponen_sound_2006}. The mean-field Hamiltonian is diagonalized by a Bogoliubov transformation with quasi-particle annihilation operators $\alpha_{k,\gamma}$ and $\beta_{k,\gamma}$
	\bea{
	H_{\text{mf}}(J,J',U,\mu)&=\sum_{k,\gamma} E_{k,\gamma} \vb{\alpha_{k,\gamma}^\dag \alpha_{k,\gamma} + \beta_{k,\gamma}^\dag \beta_{k,\gamma}}\com\label{hdiag}
	}
	where $\gamma=1,2$
	and
	\bea{
	E_{k,\gamma}&=\sqrt{\epsilon^2_{k,\gamma}+\Delta^2}\dt
	}

	For small attractive interactions $U<0$ the ground-state wave function of Eq.~\ref{meanFieldH} is the well known BCS wave function \cite{leggett_quantum_2006} at zero temperature, while at finite temperature $T$ the ground state is a density matrix, where the order parameter is given by the self-consistent gap equation
	\bea{
	\Delta &= -\frac{U}{2M} \sum_{k,\gamma} \frac{\Delta}{2E_{k,\gamma}}\tanh\vb{\frac{E_{k,\gamma}}{2k_BT}}\com
	}
	with Boltzmann constant $k_B$. 

\section{Model for the time evolution}\label{model}
	In this section we present the analytical model used to evaluate the time-dependent expectation values of observables after a sudden ramp of the lattice depth. In particular, we are interested in the momentum occupation numbers of the two bands $\gamma=1,2$
	\bea{
	P_{k,\gamma}(t)&=\ave{e^{-itH(J_f,J_f',U_f)}\, a_{k,\gamma}^\dag a_{k,\gamma} \,e^{itH(J_f,J_f',U_f)}}\label{Pkg}
	}
	and the time-dependent order parameter
	\bea{
	\Delta(t)=-\frac{U}{2M}\sum_{p,\gamma}\ave{e^{-itH}\, b_{-k,\gamma} a_{k,\gamma} \,e^{itH}}\dt
	}
	Here angle brackets denote the expectation value with respect to the (thermal) BCS state, as obtained by diagonalizing the initial mean-field Hamiltonian $H_{mf}( J_i, J_i', U_i,\mu_i)$ with initial values $J_i$, $J_i'$, $U_i$ and $\mu_i$. The parameters $J_f$, $J_f'$ and $U_f$ denote the corresponding quantities after the ramp. The chemical potential in the final Hamiltonian does not contribute to the time evolution. The reduced Planck constant $\hbar$ is set to $1$ throughout. 
	We focus on quenches that increase the lattice depth. This implies that the atom-atom interaction remains attractive. 
	Repulsive interactions are experimentally accessible through magnetic Fano-Feshbach resonances and the simultaneous
	quench of the applied magnetic field. Our derivations are also valid in this regime.

	While we use mean-field theory to determine the initial state we use the full Hamiltonian $H(J_f,J_f',U_f)$ for the time propagation going beyond mean-field theory.
	Finally, it suffices to consider the time evolution of spin-down fermions as the Hamiltonian is symmetric with respect to the exchange of spin species for equal populations.

	We first consider a ramp to a sufficiently deep lattice such that the dynamics after the ramp are determined by the interaction part of the Hamiltonian. Then the hopping part is negligible and we can assume  $J_f=J_f'=0$. Some algebra, presented in App.~\ref{timeEvolCalc}, leads to the following analytic expression for the time evolution of each momentum mode in the two bands
	\bea{
	P_{k,\gamma}(t)&=n_{k,\gamma}
	+ 2\sin\vb{t U_f} {\rm Im}\vb{G_{k,\gamma} D^*} \eqb
	\quad+ 2\vsb{1-\cos(t U_f)} Z_{k,\gamma}
	\label{tEvol}\com
	}
	with all time dependence isolated in the $\sin$ and $\cos$, the initial momentum occupation
	\bea{
	n_{k,\gamma}&=\ave{a_{k,\gamma}^\dag a_{k,\gamma}}=\frac 12 \vcb{1-\frac{\epsilon_{k,\gamma}}{E_{k,\gamma}}\tanh\vb{\frac{E_{k,\gamma}}{2k_BT}}}\com
	}
	the initial momentum-resolved pairing field
	\bea{
	G_{k,\gamma}&=\ave{b_{-k,\gamma}a_{k,\gamma}}=\frac{\Delta}{2E_{k,\gamma}} \tanh\vb{\frac{E_{k,\gamma}}{2k_BT}}
	}
	and
	\bea{
	Z_{k,\gamma}&=
	(1-n_{k,\gamma})\abss D +W_{k,\gamma}\dt\label{zk}
	}
	Finally $D=\Delta/U$ is the scaled order parameter. The quantity $W_{k,\gamma}$ is time independent and given in App.~\ref{timeEvolCalc}. 
	Note that both $\epsilon_{k,\gamma}$ and $E_{k,\gamma}$ depend on $J_i$ and $J_i'$ and $\Delta$ is computed for the initial hopping and interaction parameters.
	Lastly, ${\rm Re} (z)$ and ${\rm Im} (z)$ are the real and complex part of $z$, respectively.

	\begin{figure}[t]
	    \includegraphics[width=\linewidth]{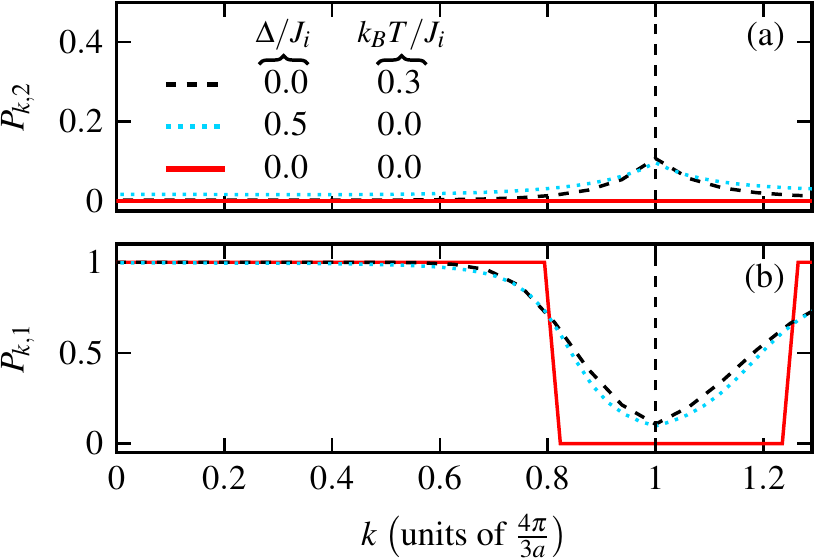}
	    \caption{(color online) 
	    Initial momentum distribution $n_{k,1}$ (panel (b)) and $n_{k,2}$ (panel (a)) along the $\Gamma$-K-M-direction in bands 1 and 2, respectively, for filling $n=0.45$. The units of $k$ are chosen such that $k=1$ at the K-point, which is situated at the edge of the first Brillouin zone, indicated by the dashed vertical line. Different curves show different values for the order parameter $\Delta$ and temperature $T$ as indicated in the legend. For all cases $J_i'=0.2J_i$.
	    }
	    \label{fig:bcsHoneyKIni}
	    \end{figure}

	\begin{figure*}[t]
	    \includegraphics[width=\textwidth]{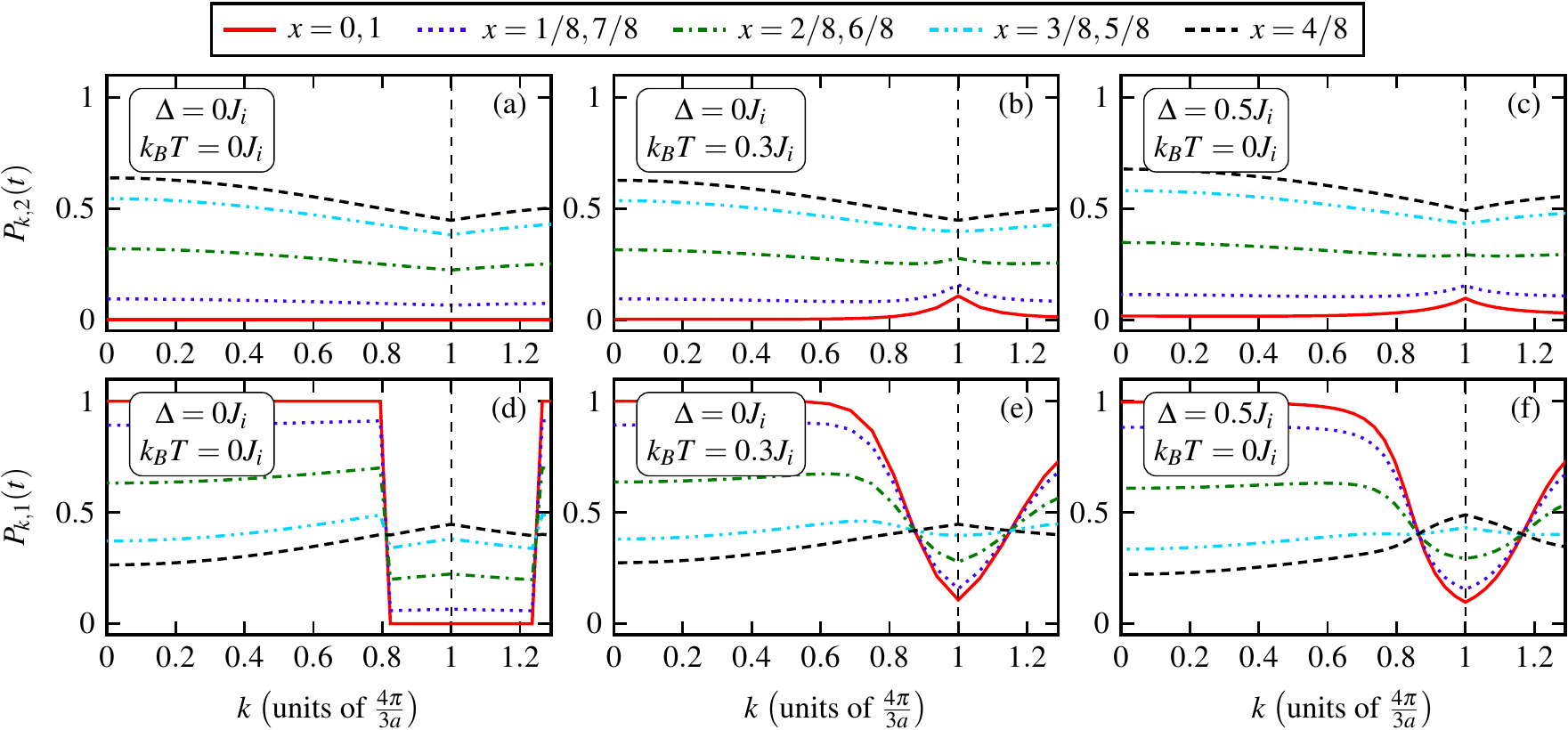}
	    \caption{(color online) 
	    Momentum distribution $P_{k,\gamma}(t)$ after a sudden ramp to a lattice with zero tunneling and filling fraction $n=0.45$. We show the momentum distribution along the $\Gamma$-K-M direction at several snapshots in time $x=t\abs{U_f}/(2\pi)$ as indicated in the legend above the figure. The units of $k$ are chosen as in Fig.~\ref{fig:bcsHoneyKIni}. The top and bottom rows show populations in the upper and lower band, respectively. The time point $t=0$ (solid red line) corresponds to the initial momentum distribution before the ramp, which is again obtained at $t=2\pi/\abs{U_f}$ due to the periodicity of the time evolution. Panels (a) and (d) show distributions at zero temperature and gap, while panels (b) and (e) show zero gap and finite temperatue $k_BT=0.3J_i$. Panels (c) and (f) show finite gap $\Delta=0.5J_i$ and zero temperature, which corresponds to $U_i=-2.68J_i$.  Finally, in all cases $J_i'=0.2J_i$, $\mu=0$, $J_f=J_f'=0$ and the results are valid for any sufficiently large $U_f$.
	    }
	    \label{fig:bcsHoneyK1}
	    \end{figure*}

	Next, we consider a quench to a final lattice depth, where small hopping parameters $J_f,J_f'\ll \abs{U_f}$ remain, and solve it perturbatively by using the Suzuki approximation for the exponential of the final Hamiltonian
	\bea{
	e^{it(H_J+H_U)}&= e^{itH_J/2} e^{itH_U} e^{itH_J/2}+\mathcal O(t^3)\label{suzuki}\com
	}
	where of course $H_J$ and $H_U$ depend on $J_f$, $J_f'$ and $U_f$. 

	As explained in App.~\ref{timeEvolCalc1} we again obtain Eq.~\ref{tEvol}, but make the replacements
	\bea{
	G_{k,\gamma}&\rightarrow \mathcal G_{k,\gamma}(t)=\exp\vb{i t\nu_{k,\gamma}}\frac{\Delta}{2E_{k,\gamma}} \tanh\vb{\frac{E_{k,\gamma}}{2k_BT}}\label{finJIk}
	}
	and
	\bea{
	D&\rightarrow \mathcal D(t)= -\frac 1{2M} \sum_{p,\gamma} \mathcal G_{p,\gamma}(t)\label{finJD}\com
	}
	which are now time dependent quantities. Here, the band energies $\nu_{k,\gamma}$ are $\epsilon_{k,\gamma}$ evaluated at $J_f$, $J_f'$ and $\mu=0$ rather than $J_i$, $J_i'$ and $\mu_i$. Note that the energies $E_{k,\gamma}$ are evaluated at $J_i$ and $J_i'$ but that $G_{p,\gamma}$ and $D$ are replaced in the definitions for $Z_{k,\gamma}$ and $W_{k,\gamma}$. 

	The accuracy of the Suzuki approximation can be estimated from the strength of terms of cubic order in time. 
	These have two contributions, one proportional to $|U_f|^2J_ft^3$ and the other to $\abs{U_f}J_f^2t^3$. In our case $J_f\ll \abs{U_f}$ and we therefore require $t^3\ll 1/(\abs{U_f}^2J_f)$.

	In summary, we have derived an expression for the time evolution of the momentum occupation, which can be evaluated analytically except for straightforward numerical summations in $W_{k,\gamma}$. While we focus on the density and pairing field, this calculation can be extended to the time evolution of other operators.

\section{Observing collective oscillations}\label{oscillations}
	\subsection{Time evolution of the momentum modes for a quench to zero hopping}\label{mom0}
	A sudden ramp of the lattice depth to a deep lattice, where $J_f=J_f'=0$, induces collective oscillations of the quasi-momentum occupation numbers $P_{k,\gamma}(t)$. In order to get an understanding for these oscillations we first investigate the initial quasi-momentum distribution, which is shown in Fig.~\ref{fig:bcsHoneyKIni}  for a filling fraction $n$ slightly less than $1/2$ and several $\Delta$ and $T$. The filling fraction $n=1/(2M)\sum_{p,\gamma} n_{p,\gamma}$ is the mean number of particles per site per spin state. 

	For non-interacting fermions at half filling $n=1/2$ and $T=0$ the lower of the two bands is completely filled. The upper band is empty. Population is removed around the Dirac ($K$-) points in the lower band for slightly smaller $n$. Finite temperature, on the other hand, transfers population to the second band, predominantly around the Dirac points. 
	A kink in the quasi-momentum profiles appears at this point, as the two bands touch linearly.
	A finite order parameter has similar effects. In fact, a comparison of the curves in Fig.~\ref{fig:bcsHoneyKIni} shows that distinguishing a paired state with finite order parameter from a finite temperature state by looking at the initial momentum distribution only, is hard if not impossible. 

	The time evolution of $P_{k,\gamma}(t)$ in Eq.~\ref{tEvol} is periodic with $k$-independent frequency $\abs{U_f}/(2\pi)$. Moreover, as $G_{k,\gamma}$ and $D$ are real for $J_f=J_f'=0$, the occupation numbers $P_{k,\gamma}(t)$  simplify to $n_{k,\gamma}+2(1-\cos(tU_f)) Z_{k,\gamma}$ and oscillate in phase.
	Figure \ref{fig:bcsHoneyK1} shows the momentum distributions at different times $t$ in the first oscillation cycle $0<t<2\pi/\abs{U_f}$ for several values of temperature and order parameter. After half an oscillation period at $t=\pi/\abs{U_f}$ we find that momentum modes with small initial occupation have high occupation and vice versa. In particular, we observe a significant occupation of the second band for all momentum modes. 

	We find numerically that the main contribution to the amplitude of the momentum oscillations, $Z_{k,\gamma}$, does not depend on the order parameter $\Delta$. Out of the terms that do depend on $\Delta$ the dominant one is the first term in Eq.~\ref{zk}, $(1-n_{k,\gamma})\abss \Delta/U^2$. It enhances the population of momentum modes with small initial occupation $n_{k,\gamma}$ at $t=\pi/\abs{U_f}$. We observe this when comparing $P_{k,\gamma}(t=\pi/\abs{U_f})$ (dashed black line) of a state with a finite order parameter in Figs.~\ref{fig:bcsHoneyK1}(c) and (f) with that of a finite temperature state in Figs.~\ref{fig:bcsHoneyK1}(b) and (e). In fact, the enhancement occurs for all momentum modes in the upper band as well as for those close to the Dirac point in the lower band.  

	We conclude that the oscillation frequency of the momentum modes after a sudden ramp of the lattice depth to $J_f=J_f'=0$ is a direct measure for the interaction strength between atoms. Furthermore, we find small differences in the time evolution of momentum modes between finite-gap and finite-temperature states. The kink of the momentum distribution at the $K$-point of our hexagonal lattice remains observable after the ramp. Measuring both effects in experiment may, however, be limited by the current resolution of time of flight images. 
	\begin{figure*}[t]
	    \includegraphics[width=\textwidth]{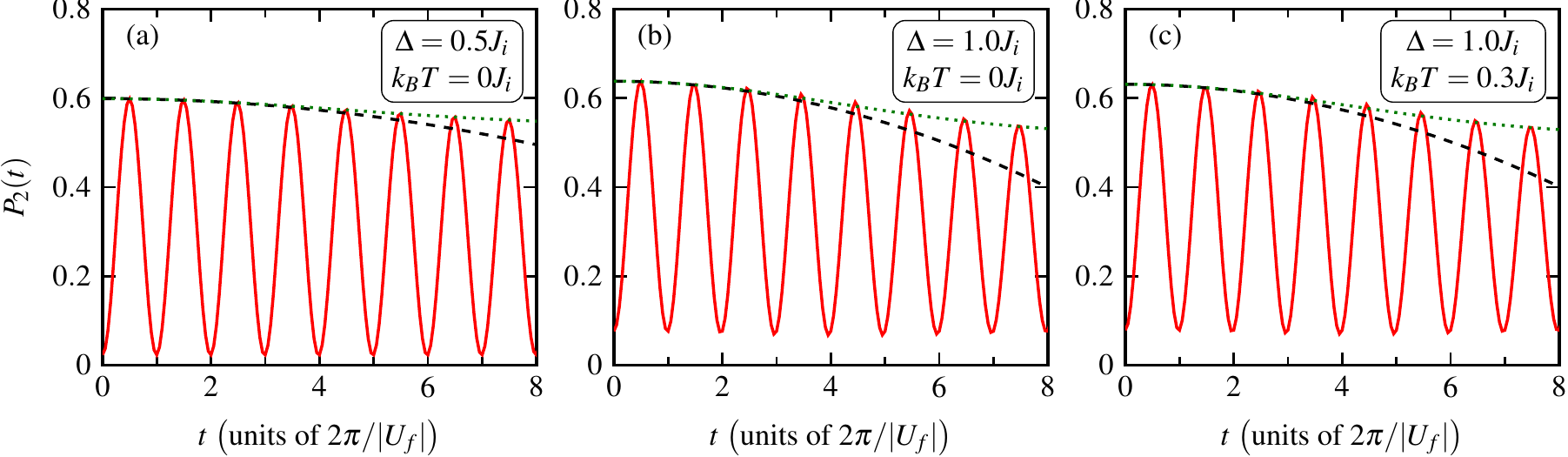}
	    \caption{(color online) Total occupation of the second band $P_2(t)=1/M \sum_q P_{q,2}(t)$ (solid red line) as a function of time with a residual finite hopping after the ramp and filling fraction $n=0.45$. The dotted green line shows the envelope of the oscillations $\mathcal P_2(t)$ as defined in Eq.~\ref{halfCyc}. For small times it is well approximated by a quadratic time dependence (dashed black line). For all panels the final hopping strength $J_f=0.02U_f$, the ratios $J_{f}'/J_f=J_i'/J_i=0.2$ and the results are valid for any sufficiently large $U_f$. Panels (a) and (b) show zero temperature data with $\Delta=0.5J_i$ (corresponding to $U_i=-2.68J_i$) and $\Delta=1J_i$ ($U_i=-3.45J_i$), respectively. Panel (c) shows data for $\Delta=1.0J_i$ and $k_BT=0.3J_i$, implying $U_i=-3.53J_i$.
	    }
	    \label{fig:finHopp}
	    \end{figure*}	

	\begin{figure}
	    \includegraphics[width=\linewidth]{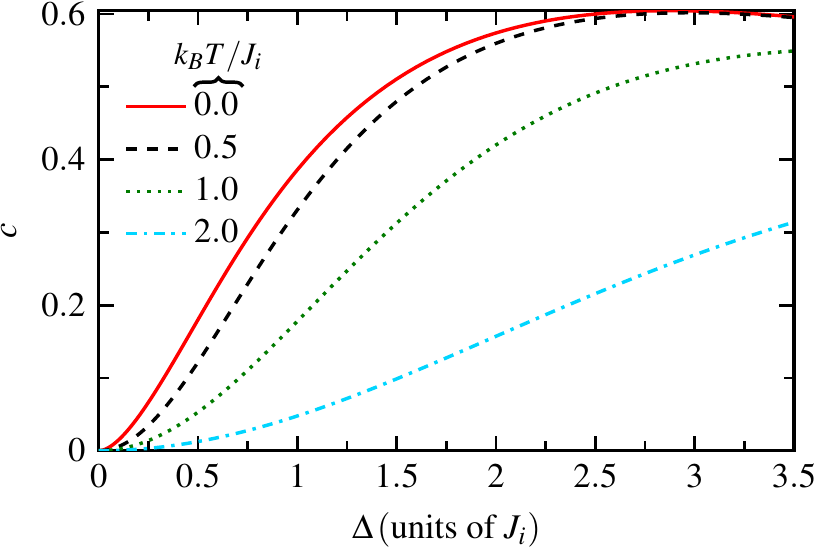}
	    \caption{(color online) Quadratic coefficient $c$, quantifying the damping of the oscillations of the second-band occupation $P_2(t)$, as a function of order parameter $\Delta$ for constant filling fraction $n=0.45$ and $J_f'=0.2J_f$. Different curves show $c$ at different temperatures as indicated in the legend. A larger coefficient $c$ indicates a faster damping.
	    }
	    \label{fig:finHopp2}
	    \end{figure}

	\subsection{Time evolution of the momentum modes for a quench to small finite hopping}\label{momfin}
	Even for ramps to deep lattices, hopping between lattice sites will not be completely negligible. We take this into account perturbatively in Eq.~\ref{tEvol} with the definitions from Eqs.~\ref{finJIk} and \ref{finJD}. Most notable we find dephasing of the momentum occupation numbers for an initial state with finite order parameter. The pairing fields $\mathcal G_{p,\gamma}(t)$ then evolve with a different frequency for each momentum and band index. This causes the summands in $\mathcal D(t)$ to dephase and eventually leads to damping of the oscillations of the occupation numbers. 

	The damping is illustrated in Fig.~\ref{fig:finHopp}, where we show the total population of the second band $P_{2}(t)=2n-P_1(t)=1/M \sum_k P_{k,2}(t)$. We observe that $P_2(t)$ is close to a minimum, whenever $t$ is a multiple of $2\pi/\abs{U_f}$. In fact, at these time points the quantitites $\sin(tU_f)$ and $1-\cos(tU_f)$ in Eq.~\ref{tEvol} are zero and  the occupation numbers $P_{k,\gamma}(t)$ are equal to their initial values $n_{k,\gamma}$. The most pronounced dephasing effects can be observed, when $P_{2}(t)$ is close to a maximum, half way in between two such revivals at $t=t_j=(j-1/2)2\pi/\abs{U_f}$, with positive integer $j$. In Fig.~\ref{fig:finHopp} we see that the dephasing of momentum modes causes $P_{2}(t_j)$ to decrease over several time-evolution cycles. Comparing Figs.~\ref{fig:finHopp}(a) and (b) it is furthermore evident that $P_{2}(t_j)$ decreases more rapidly for larger values of $\Delta$.

	This motivates a closer investigation of $P_2(t)$. For this purpose it is useful to define the envelope of the occupation numbers
	\bea{
	\mathcal P_{2}(t)=\sum_k \vsb{n_{k,2}+4Z_{k,2}(t)}\label{halfCyc}\com
	}
	which is obtained by evaluating the periodic quantities $\sin(t_jU_f)=0$ and $1-\cos(t_jU_f)=2$ at $t=t_j$ in Eq.~\ref{tEvol}, but keeping the time dependence of $\mathcal G_{k,\gamma}(t)$, $\mathcal D(t)$ and $Z_{k,\gamma}(t)$. Figure \ref{fig:finHopp} shows that $\mathcal P_2(t)$ closely follows the maxima of $P_2(t)$. It has a quadratic time dependence for small $tJ_f$ and $tJ_f'$
	\bea{
	\mathcal P_2(t)&=\mathcal P_{2}(0)\vsb{1-c(\Delta) J_f^2 t^2 + \mathcal O ([J_ft]^3)}\com\label{upEnvelope}
	}
	where $c$ is a function of the order parameter $\Delta$, the filling fraction $n$ and the relative hopping strength $J_f'/J_f$. Nevertheless, we only make the $\Delta$ dependence explicit as we expect the other two quantities to be approximately constant in experiments.
	This dependence is analytically confirmed by evaluating $P_{k,\gamma}(t)$ using the Lie first-order approximation for the Hamiltonian
	\bea{
	\exp\vb{itH(J_f,U_f)}&=\exp\vb{itH_J} \exp\vb{itH_U} +\mathcal O(t^2)\com
	}
	instead of the Suzuki second-order approximation from Eq.~\ref{suzuki}. We obtain the same time evolution as for $J_f=J_f'=0$, because the hopping part of the Hamiltonian $H_J$ commutes with the observable $a_{k,\gamma}^\dag a_{k,\gamma}$. Therefore $P_{k,\gamma}(t)$ and $P_2(t)$ are independent of $J_f$ and $J_f'$ and hence $\mathcal P_2(t)$ does not have a contribution linear in time.

	The quadratic approximation for $\mathcal P_2(t)$ agrees well with the exact $\mathcal P_2(t)$ for the first few oscillations. Afterwards terms of cubic and higher order in time are important. We, however, do not expect the Suzuki approximation in Eq.~\ref{suzuki} to be valid in that regime. An estimate for its validity is given by the condition $t\abs{U_f}/(2\pi)\ll (\abs{U_f}/J_f)^{1/3}$, where the right hand side equals $3.7$ for $J_f=0.02\abs{U_f}$, as used throughout this paper.

	In Fig.~\ref{fig:finHopp2} we plot the quadratic coefficient $c(\Delta)$ obtained from the analytic expansion of $\mathcal P_2(t)$ as a function of $\Delta$ for several temperatures.
	It vanishes for a zero order parameter, since Eqs.~\ref{finJIk} and \ref{finJD} vanish ($\mathcal G_{k,\gamma}(t)=\mathcal D(t)=0$) and the envelope $\mathcal P_2(t)$ is independent of time. In other words the oscillations do not damp, when the initial state is a non-interacting Fermi gas. The coefficient $c(\Delta$) increases quadratically for $\Delta\ll J_i$ and reaches a maximum for larger values of the order parameter. Finally, the damping coefficient decreases for increasing temperature. This motivates the use of $c(\Delta)$ to detect the order parameter experimentally.

	In summary we see that the BCS type correlations lead to an increased dephasing of the different momentum modes, which leads to additional damping of the oscillations. Isolating this effect from other damping origins may, however, be challenging in experiment.

	\begin{figure*}[t]
	    \includegraphics[width=\textwidth]{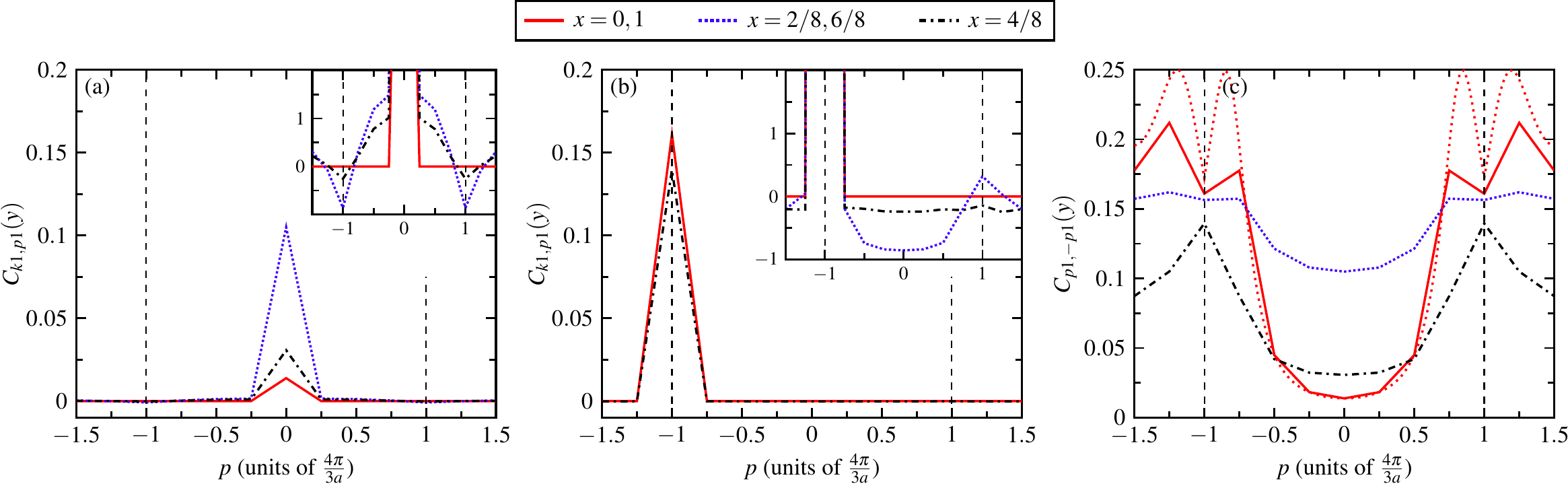}
	    \caption{(color online) Density-density correlations in momentum space $C_{k\gamma,p\sigma}(t)$ in the first band ($\gamma=\sigma=1$) after a sudden ramp of the lattice depth  in a periodic honecycomb lattice with $12 \times 12$ unit cells at $k_BT=0.3J_i$, $\Delta=1J_i$ ($U_i=-3.53J_i$)  and filling fraction $n=0.45$. All panels show several snapshots in dimensionless time $x=t\abs{U_f}/(2\pi)$, where the coloring scheme determines the time point as indicated in the legend.
	    We use $\mbf k = (0,0)^T$ and the Dirac point ($\mbf k=1/3(\mbf k_1+\mbf k_2)$) in panels (a) and (b), respectively, and show the correlations as a function of p for $\mbf p = (p,p)^T$. The insets show a zoom of the full figure with the y-axis scaled by a factor of $1000$. 
	    Panel (c) shows the correlations within the first band for $\mbf p=-\mbf k$ and $\mbf p=(p,p)^T$. The dotted red line shows $|I_p(t)|^2$ computed for $102 \times 102$ lattice sites for comparison.
	    In all panels the units of $p$ are chosen such that $p=-1,1$ at the $K$-points, which are indicated by dashed vertical lines.
	    }
	    \label{fig:honDdC1}
	    \end{figure*}

	\subsection{Time evolution of the order parameter}\label{gap}
	The time evolution of the pairing order parameter $\Delta(t)$ after a quench has been simulated extensively \cite{volkov_collisionless_1973,barankov_collective_2004,tomadin_nonequilibrium_2008,warner_quench_2005,yuzbashyan_nonequilibrium_2005,yuzbashyan_solution_2005,barankov_synchronization_2006,dzero_spectroscopic_2007,andreev_nonequilibrium_2004,szymanska_dynamics_2005,yuzbashyan_dynamical_2006,yuzbashyan_relaxation_2006,scott_rapid_2012,yuzbashyan_quantum_2015}. Here we present our results for $\Delta(t)$ and compare with Ref.~\cite{scott_rapid_2012}, which we found to be most closely related to our calculations. We evaluate $\Delta(t)$ using the formalism introduced in Sec.~\ref{model} and find for a sudden ramp to a lattice with $J_f=J_f'=0$
	\bea{
	\Delta(t)&=\Delta \exp\vb{itU_f}\com\label{tEvolGap}
	}
	independent of temperature. Hence, its amplitude is constant while its phase oscillates with the same frequency $\abs{U_f}/(2\pi)$ as the momentum modes. (In fact, we can show that for any state only the phase of the pairing order parameter oscillates in time.)

	Reference \cite{scott_rapid_2012} solves the mean-field Bogoliubov-de Gennes (BdG) equations for a homogenous system at zero temperature for either slow or fast changes of the interaction strength. Unlike our simulations, they assume that the system remains in a BCS state for all times.
	For both fast and slow ramps, they find damped oscillations of the amplitude of the order parameter around an average value $\Delta_\infty$, with a frequency of $2\abs{\Delta_\infty}/(2\pi)$. Only for ramps slow compared to their Fermi energy the average value $\Delta_\infty$ equals the order parameter of the BCS ground state of the final Hamiltonian.
	In other words, our and the BdG models make different predictions for the oscillation frequency as well as the average value.

	We note that Eq.~\ref{tEvolGap} is valid for an infinitely fast ramp of the lattice depth and found to be true for one dimensional, square and honeycomb lattices. In contrast, we expect that the BdG simulations of Ref.~\cite{scott_rapid_2012} are only valid for slow quenches to interaction strengths, that are not too large and do not lead to high energy excitations. For fast quenches, on the other hand, we trust our calculations. In summary, even though the two simulations are similar in spirit, they are complementing each other by exploring different quench regimes.

	\subsection{Time evolution of the density-density correlation function}\label{highOrder}

	The density-density correlations of the BCS ground state
	\bea{
	C_{k\gamma,p\sigma}&=\ave{a_{k,\gamma}^\dagger a_{k,\gamma} b_{p,\sigma}^\dag b_{p,\sigma}} - \ave{a_{k,\gamma}^\dagger a_{k,\gamma}} \ave{b_{p,\sigma}^\dag b_{p,\sigma}}\label{ddCor1}
	}
	have been of much interest as they can be measured in experiment and are, within mean-field theory, directly proportional to the gap $C_{k\gamma,p\sigma}=\delta_{k,-p}\delta_{\gamma,\sigma} \, \abss \Delta/(4E_{k,\gamma}^2)$\cite{altman_probing_2004,carusotto_coherence_2004, greiner_probing_2005, lamacraft_particle_2006,belzig_density_2007, paananen_noise_2008,kudla_pairing_2015}.

	We compute the time-evolved density-density correlations $C_{k\gamma,p\sigma}(t)$ by inserting the exponentials $e^{itH(J_f,J_f',U_f)}$ inside all expectation values in Eq.~\ref{ddCor1}. Here, only the case of a deep lattice, such that $J_f=J_f'=0$, is considered. In principle, we could use the formalism introduced in Sec.~\ref{model} to compute $C_{k\gamma,p\sigma}(t)$. We would, however, have to evaluate an expectation value of $12$ operators. This correpsonds to $6!=720$ different terms and is therefore tedious to compute by hand. We therefore use a different approach, which, while giving less insight, is much easier to automate for higher order correlation functions. First, we separate the time dependence from the expectation values by using an identity similar to Eq.~\ref{sepTime}. Then we evaluate the time-independent expectation values of operators in lattice space instead of momentum space. This has the advantage that the expectation values of $L$ operators factor into a product of two-operator expectation values and  we do not get multi-dimensional sums as in Eq.~\ref{SumQ}. In fact, Wicks theorem can be applied \cite{ballentine_quantum_1998} and we find 
	\bea{
	\ave{c_1^\dag c_2^\dag \dots c_L^\dag c_{L'} \dots c_{2'}c_{1'}}&=\sum_{s\in S(L)} \vsb{ {\rm sign}(s) \prod_{j=1}^L \ave{c_j^\dag c_{s(j)'}}}\com\label{ddCor2}
	}
	where each of the number indices $i$ denotes a multi-index with unit-cell index $n_i$, sublattice site $C_i$ and spin $\sigma_i$. Primed indices denote a set of different independent multi-indices. Furthermore the operators $c_i=a_{n_i,C_i}$ for $\sigma_i=\down$ and $c_i=b^\dag_{n_i,C_i}$ for $\sigma_i=\up$. Finally, $S(L)$ is the set of all permutations of the numbers $1,2,\dots,L$ and ${\rm sign}(s)$ denotes the sign of the permutation $s$.
	Note that it is important that the left hand side of Eq.~\ref{ddCor2} is normal ordered in the sense that all $c_i^\dag$ operators are left of the $c_j$ operators. 

	\begin{figure*}[t!]
        \includegraphics[width=\textwidth]{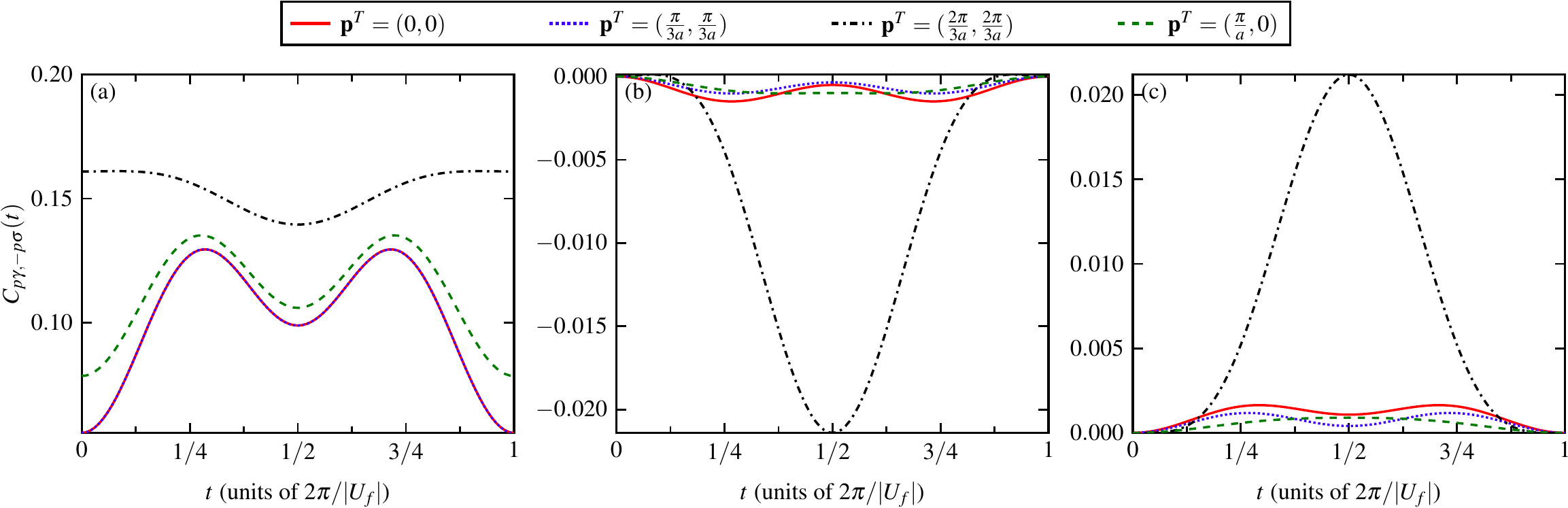}
        \caption{(color online) Density-density correlations in momentum space $C_{p\gamma,-p\sigma}(t)$ after a sudden ramp of the lattice depth  in a periodic honeycomb lattice with $12 \times 12$ unit cells at $k_BT=0.3J_i$, $\Delta=1J_i$ ($U_i=-3.53J_i$)  and filling fraction $n=0.45$. We show the correlations as a function of time for several momentum values $\mbf p$, where the coloring scheme determines the momentum as indicated in the legend. Panel (a) shows the correlations within the first band, panel (b) the correlations between the first and the second band and (c) shows  $C_{p1,-p1}(t)-\abss{I_{p1}(t)}$ within the first band. The latter expression is zero when calculated within mean-field theory. Finally, we note that the red and dotted blue lines overlap in panel (a).}
        \label{fig:honDdC2}
        \end{figure*}

	The density-density correlations are periodic with the same frequency $\abs{U_f}/(2\pi)$ as the momentum occupation numbers. In fact, $C_{k\gamma,p\sigma}(t)= \mathcal C_1 + \mathcal C_2 \cos(tU_f) + \mathcal C_3 \cos^2(tU_f)$ with time-independent, but momentum- and band-dependent, real coefficients $\mathcal C_i$, $i=1,2,3$.
	Furthermore Fig.~\ref{fig:honDdC1} (a) and (b) show that throughout the whole time evolution the correlations are dominated by momenta $\mbf p = -\mbf k$. A finite background remains with values about a factor of $100$ smaller. While we show results for $k_BT=0.3J_i$, we note that the results for different temperatures and in particular zero temperature are qualitatively the same. 

	A closer investigation of the $\mbf p=-\mbf k$ correlations in Fig.~\ref{fig:honDdC1} (c) reveals that the amplitude of the oscillations at the Dirac point is smaller than at any other momentum point. The same holds for the correlations within the second band as we see in Fig.~\ref{fig:honDdC2} (a). In contrast, Fig.~\ref{fig:honDdC2} (b) shows that the opposite is true for the correlations between the first and the second band. In fact, these two bands develop significant anti-correlations, i.e. negative values of the correlation function, throughout the time evolution.

	Figure \ref{fig:honDdC2} (c) compares results obtained within mean-field theory with our exact results. In fact, when using the mean-field Hamiltonian for the time-evolution the initial state remains a BCS type state and for all times and $\ave{a_{k,\gamma}^\dagger a_{k,\gamma} b_{p,\sigma}^\dag b_{p,\sigma}} = \ave{a_{k,\gamma}^\dagger a_{k,\gamma}} \ave{b_{p,\sigma}^\dag b_{p,\sigma}}+ \ave{a_{k,\gamma}^\dag b_{p,\sigma}^\dag} \ave{b_{p,\sigma} a_{k,\gamma}}$. Hence, this implies that the difference
	\bea{
	C_{k1,p1}(t)-\delta_{k,-p}\abss{I_{k1}(t)}\label{denDen1}
	}
	is strictly zero, where $I_{k,1}(t)=\ave{e^{-itH(J_f,J_f',U_f)}b_{-k,1}a_{k,1}e^{itH(J_f,J_f',U_f)}}$. 
	In other words, a mean-field theory predicts a zero background in the insets of Figs.~\ref{fig:honDdC1}(a) and (b), where we find small non-zero values from the exact $J_f=0$ simulations. Furthermore Fig.~\ref{fig:honDdC2} (c) shows the expression in Eq.~\ref{denDen1}, at $\mbf p=-\mbf k$ evaluated within our exact theory. We see that the correlations at $\mbf p=-\mbf k$ have small deviations from the mean-field theory for all momenta, but which are particularly pronounced at the Dirac point. 

\section{Conclusions and Outlook}\label{sec:summary}
	We have analyzed the exact time evolution of a BCS state after a sudden quench of the lattice depth. For zero tunneling after the quench we find undamped collective oscillations of the momentum occupation numbers with frequency $\abs{U_f}/(2\pi)$. The observation of these oscillations is experimentally accessible through time-of-flight measurements. Small finite hopping after the quench leads to dephasing of different momentum modes and a corresponding damping of the oscillations. On short time scales we observe that at any fixed temperature the damping is stronger for larger order paramter $\Delta$. In particular, our perturbative calculations find no damping at all if the initial state is a non-interacting Fermi gas. Measuring the quadratic damping coefficient may therefore be used to estimate the size of the order parameter. 

    We note, however, that the measurement of the dephasing time will be challenging and always only be an indirect proof of a finite order parameter for the fermions. For example, additional numerical calculations for small lattice sizes, presented in App.~\ref{pureNumeric}, show that an improved description of the initial thermal-equilibrium state leads to additional damping. A direct comparison of the analytical and the small size numerical model has to be taken with care due to the significant difference in lattice size and topology. Still, it may be challenging to experimentally distinguish different damping mechanisms.


	Finally, experimental limitations might make it hard to extract the dephasing time. In order to mitigate the effect of additional dephasing mechanisms the contribution of the BCS-type correlations to the damping of the oscillations can be increased by using larger hopping after the quench, as can be seen from Eq.~\ref{upEnvelope}. Although our perturbative results are not valid in that regime, we expect that the qualitative behaviour remains the same. In experiments additional dephasing can occur due to density inhomogeneities in the initial state. We expect this to lead to small corrections as mass transport is absent when $J=0$ and very small for small non-zero $J$. Based on findings with similar quench experiments with ultra-cold bosonic atoms \cite{tiesinga_collapse_2011,buchhold_creating_2011,will_coherent_2011} it  will be more important to include the effect of weak confining potentials after the quench. A spatially varying on-site energy leads to  additional dephasing. The experiments with bosons have shown that confinement effects can to a large extent be  mitigated, for example by using shallow traps or box potentials \cite{gaunt_bose-einstein_2013,corman_quench-induced_2014}. Similar observations may be expected for Fermions, which makes the investigation of confinement effects an interesting direction for future research. 

    For the time evolution of the order parameter we find oscillations of the phase with frequency $\abs{U_f}/(2\pi)$. This differs from previous results \cite{volkov_collisionless_1973,barankov_collective_2004,tomadin_nonequilibrium_2008,warner_quench_2005,yuzbashyan_nonequilibrium_2005,yuzbashyan_solution_2005,barankov_synchronization_2006,dzero_spectroscopic_2007,andreev_nonequilibrium_2004,szymanska_dynamics_2005,yuzbashyan_dynamical_2006,yuzbashyan_relaxation_2006,scott_rapid_2012} obtained from treating both the initial state as well as the time evolution within the mean-field approximation. Our results, are valid for ramps fast compared to the timescale of interactions, while we expect mean-field theory to be valid in the opposite limit. 
    Also we note that Ref.~\cite{scott_rapid_2012}, which we found to be most closely related to our work, considers a continuous system, while ours is a discrete lattice. Although it is not clear how to take the continuum limit, the fact that we observe qualitatively similar time evolutions for different discrete lattice topologies suggests that the comparison to a continous system is valid.
    Still, the two approaches complement each other by exploring different quench regimes.


	The lowest two bands of the honeycomb lattice touch linearly at the Dirac point. This gives rise to a kink in the momentum distribution, which remains visible throughout the time evolution.
	We further observe that the density-density correlations, which perform periodic oscillations with the same frequency $\abs{U_f}/(2\pi)$ as the momentum occupation numbers, show pronounced differences in the amplitude of the oscillations at the Dirac point as compared to other momenta.
	Both within the first and second band the oscillation amplitude is significantly smaller at the Dirac point. The opposite is true for the correlations between the first and the second band. While initially uncorrelated, the system develops strong anti-correlations between those two bands at the Dirac point.

	\begin{acknowledgments}
	This work has partially been supported by the National Science Foundation Grant No.~PHY-1506343. M.N. and L.M. acknowledge support from the Deutsche Forschungsgemeinschaft through the SFB 925, L.M. acknowledges support from  the Hamburg Centre for Ultrafast Imaging, and from the Landesexzellenzinitiative Hamburg, supported by the Joachim Herz Stiftung. M.N. acknowledges support from the German Economy Foundation.
	\end{acknowledgments}

\begin{appendix}

	\section{Detailed calculation for the time evolution procedure}\label{timeEvolCalc}
	Here we present details for the calculation of the time evolution expression in Eq.~\ref{tEvol}. The calculation is most elegant when evaluating parts of the expression in momentum space and others in lattice space. Therefore it will be convenient to write the Hamiltonian of Eqs.~\ref{eq:Hamiltonian}-\ref{eq:HamMomChemPot} in lattice space
	\bea{
	H_J&=-J \sum_{\nn{ nC,mD }} \vb{a_{m,D}^\dagger a_{n,C}+b_{m,D}^\dagger b_{n,C}} \eqb
	-J' \sum_{ \nnn{nC,mC}} \vb{a_{m,C}^\dagger a_{n,C}+b_{m,C}^\dagger b_{n,C}} \label{eq:lattice_space_hamiltonian_j}\\
	H_U&=U\sum_{n,C} a_{n,C}^\dagger a_{n,C} b_{n,C}^\dagger b_{n,C}\label{eq:lattice_space_hamiltonian_U}\\
	H_\mu&=-\mu \sum_{n,C}\vb{a_{n,C}^\dagger a_{n,C}+ b_{nC}^\dagger b_{n,C}}\label{eq:lattice_space_hamiltonian_mu}\com
	}
	where $\nn{nC,mD}$ denotes sums over nearest neighbours, while $\nnn{nC,mD}$ denotes sums over next-nearest neighbours. The operators $a_{nC}^\dag$ ($b_{nC}^\dag$) and $a_{nC}$ ($b_{nC}$) create and annihilate a spin down (up) fermion in the unit cell $n$ with sublattice site $C=A,B$ and are related to the momentum space operators through the site-specific Fourier transformations
	\bea{
	a_{k,C}&=\frac 1{\sqrt M}\sum_n e^{-i\mbf k\cdot \mbf n} a_{n,C}\com\label{fourier}
	}
	where $\mbf n$ is the vector pointing to the origin of the $n$-th unit cell.
	Equivalent Fourier transforms are defined for the $b$ operators.

	\subsection{Zero hopping}\label{timeEvolCalc0}
	It is instructive to begin with the calculation of Eq.~\ref{Pkg} for the $J_f=J_f'=0$ case.
	The simple form of $H_U$ in lattice space is exploited by expanding $a_{k,\gamma}^\dag a_{k,\gamma}$ in terms of the  operators $a_{n,C}^\dagger a_{m,D}$, with $C,D=A,B$. The time evolution operator $\exp\vb{itH_U}$ is readily applied to each of the terms in the expansion separately
	\bea{
	e^{-itH_U}{{a_{n,C}^\dagger a_{m,D}}} e^{itH_U}&=
	{a_{n,C}^\dagger a_{m,D}}
	\bigg[ 
	\vb{1+ b_{n,C}^\dagger b_{n,C}  \vb{e^{-itU}-1}}\eqb
	\vb{1+ b_{m,D}^\dagger b_{m,D}  \vb{e^{itU}-1}}
	\bigg]\dt\label{sepTime}
	}
	By inserting this into Eq.~\ref{Pkg} and transforming all operators back into momentum space we obtain a sum of expectation values, where each term has at most six creation or annihilation operators. The expectation values are evaluated by using the Bogoliubov transformation to a non-interacting Hamiltonian (see Eq.~\ref{hdiag}) and noting that Wicks theorem is applicable to the Bogoliubov operators \cite{ballentine_quantum_1998}. For example
	\bea{
	\ave{a_{k,A}^\dag a_{k,A}}&= \frac 12 (n_{k,\alpha} + n_{k,\beta})\\
	\ave{b_{-k,A} a_{k,A}}&= \frac 12 (G_{k,\alpha} + G_{k,\beta})\dt
	}
	The result is Eq.~\ref{tEvol} with the definitions
	\bea{
	W_{k,\gamma}&=n_{k,\gamma}\vb{n^2-n}
	+(1-2n) {\rm Re}\vb{ G_{k,\gamma} D^*}\eqb
	+n^2
	-Q(k)-R(k)-(-1)^\gamma S(k)-(-1)^\gamma T(k)
	}
	and
	\bea{
	Q(k)&=\frac 1{M^2}\sum_{pq} n_{k+q-p,+} n_{p,+} n_{q,+}\label{SumQ}\\
	R(k)&=\frac 1{M^2}\sum_{pq}n_{p+q-k,+}{\rm Re}\vb{G_{p,+}^* G_{q,+}}\\
	S(k)&=\frac 1{M^2}\sum_{pq} 
	\cos\vb{\phi_k-\phi_{k+q-p}-\phi_{p}+\phi_{q}}\eqb\qquad \times n_{k+q-p,-} n_{p,-}n_{q,-}\\
	T(k)&=\frac 1{M^2}\sum_{pq} 
	\cos\vb{\phi_k-\phi_{p}-\phi_{q}+\phi_{p+q-k}}\eqb\qquad \times n_{p+q-k,-} {\rm Re}\vb{G_{p,-}^*G_{q,-}}\dt
	}
	It is furthermore convenient to define $n_{p,\pm}=\frac 12 (n_{p,2}\pm n_{p,1})$, $G_{p,\pm}=\frac 12 (G_{p,2}\pm G_{p,1})$ and the spin-independent filling fraction
	\bea{
	n&=\frac 1{2M} \sum_{p,\gamma} \ave{a_{p,\gamma}^\dag a_{p,\gamma}}=\frac 1{2M} \sum_{p,\gamma} n_{p,\gamma}\com\label{n}
	}
	which is the average number of atoms per site per spin. The remaining summations in Eq.~\ref{SumQ}-\ref{n} are evaluated numerically for equal numbers of sites $M_1$ and $M_2$ along the $\mbf e_1$ and $\mbf e_2$ directions. In fact, we choose $M_1=M_2=102$ in Figs.~\ref{fig:bcsHoneyKIni} and \ref{fig:bcsHoneyK1} and $M_1=M_2=48$ in Figs.~\ref{fig:finHopp} and \ref{fig:finHopp2}. 
	In both cases we have checked that including more lattice sites does not change the results of the calculation.

	\subsection{Small but finite hopping}\label{timeEvolCalc1}
	We now consider the time-evolution expression in Eq.~\ref{Pkg} within the Suzuki approximation (see Eq.~\ref{suzuki}). As $H_J$ commutes with $a_{k,\gamma}^\dag \, a_{k,\gamma}$ the time evolution expression immediately simplifies to
	\bea{
	P_{k,\gamma}(t)&=\ave{e^{-itH_J/2}e^{-itH_U}\, a_{k,\gamma}^\dag  a_{k,\gamma} \, e^{itH_U}e^{itH_J/2}}\label{finTEvol}\dt
	}
	Next we insert the identity $1=e^{itH_J/2}e^{-itH_J/2}$ in between all creation and annihilation operators of Eq.~\ref{finTEvol} and compute
	\bea{
	e^{-itH_J/2}a_{k,\gamma}e^{itH_J/2}&=e^{it\nu_{k,\gamma}/2} a_{k,\gamma}\label{hoppPhase}\com
	}
	where $\nu_{k,\gamma}$ is the same as $\epsilon_{k,\gamma}$, but now evaluated at $J_f$ and $J_f'$. From Eq.~\ref{hoppPhase} we see that the hopping part of the Hamiltonian simply multiplies each of the operators with a phase. By evaluating the expectation values in Eq.~\ref{finTEvol} in the same way as in Sec.~\ref{timeEvolCalc0}
	we obtain Eq.~\ref{tEvol} with the definitions from Eqs.~\ref{finJIk} and \ref{finJD}.
	
	\section{Time evolution of small systems using exact diagonalization}\label{pureNumeric}
		\begin{figure}[b]
			\includegraphics[width=\linewidth]{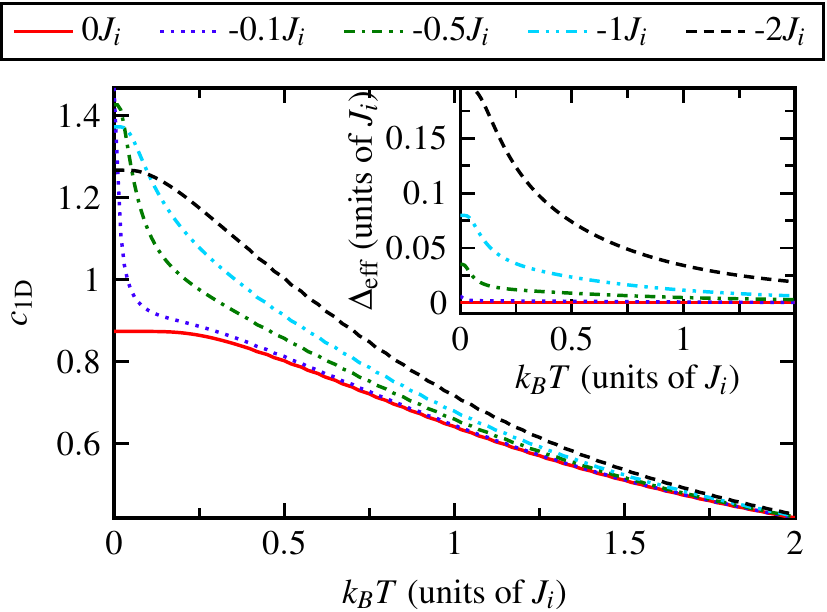}
			\caption{(color online) Quadratic coefficient $c_{\rm 1D}$, quantifying the damping of the oscillations of the momentum occupation $P_{k=0}(t)$, as a function of temperature for $U_f=-5J_i$, $J_f=0.02\abs{U_f}$ and a filling fraction of 1/3. Different curves show $c_{\rm 1D}$ for different values of $U_i$ as indicated in the legend above the figure. We use exact diagonalization with 6 lattice sites for the calculation and obtain the coefficient $c_{\rm 1D}$ from a parabolic fit to $P_{k=0}(t)$ at $t|U_f|/(2\pi)=0,1,2$. 
			The inset shows the real-space pair correlation $\Delta_{\rm eff}=-U_i/(M^2)\sum_{nm} \langle a_n^\dag a_m b_n^\dag b_m \rangle$ for the same set of parameters. The pair correlation is an estimate for the mean-field order parameter $\Delta$.}
			\label{fig:smallNumericalDamping}
			\end{figure}	    	

		\subsection{Methods}
			We extend our study to initial states with zero order parameter when simultaneously the interaction strength is non-zero. These equilibrium states of the Fermi-Hubbard Hamiltonian occur for initial temperatures higher than the critical temperature of the BCS phase transition. Calculating the subsequent time evolution falls outside the applicability of our analytical model.    We have therefore performed numerical calculations for small systems with six lattice sites and either two or three spin-up and spin-down fermions. We use a range of temperatures $0<k_BT<10J_i$ and tight binding parameters from Fig.~4.
	    
	        These numerical calculations are based on exact diagonalization of the lattice-space Hamiltonian, introduced in App.~\ref{timeEvolCalc}. In the following we briefly describe the procedure. First, we determine the matrix form of the initial Hamiltonian $H_i=H(J_i,J_i',U_i)$ in a complete set of Fock basis functions with fixed and equal number of spin-up and spin-down fermions. Next, we numerically diagonalize $H_i$ obtaining eigenvalues $E^{(i)}$ and eigenfunctions $|\psi^{(i)}\rangle$. Expectation values of an observable $\mathcal O$ with respect to initial states in thermal equilibrium at temperature $T$ are given by
	        \begin{align}
	            \ave{\mathcal O}&=\frac{1}{Z}\sum_l \exp\vb{-E^{(i)}_l/(k_BT)} \ave{\psi^{(i)}_l|\mathcal O |\psi^{(i)}_l} \label{eq:observable_time_evolution}\com\\
	            \text{where}\quad Z&=\sum_l \exp\vb{-E^{(i)}_l/(k_BT)}\nonumber
	        \end{align}
	        and $l$ is an index runnning over all eigenstates. The eigenvalues and the eigenfunctions of the final Hamiltonian are computed in a similar fashion. The time evolution of the initial states can then be expressed in terms of the overlap with the final eigenstates.
	    \subsection{Numerical results}
	    	We compute the time evolution of the momentum occupation numbers $P_{k}(t)$. All momentum modes perform collective oscillations with frequency $|U_f|/(2\pi)$. The oscillations are undamped for $J_f=0$ and we obtain a finite amount of damping that is quadratic to lowest order in time for non-zero $J_f$. Hence, these results are in good agreement with our analytical calculation and motivate a comparison of the damping strength between the two approaches. 

	    	In analogy to $\mathcal P_2(t)$ from Eq.~\ref{halfCyc} we define $\mathcal P_{k=0}(t)$ as the envelope of $P_{k=0}(t)$. We obtain $\mathcal P_{k=0}(t)$ from a quadratic fit to $P_{k=0}(t)$ at the three time points $t|U_f|/(2\pi)=0,1,2$. To good approximation these points correspond to the maxima of $P_{k=0}(t)$. As there is no contribution linear in time
	    	\begin{align}
	    		\mathcal P_{k=0}(t)=\mathcal P_{k=0}(0)\vsb{1-c_{\rm 1D}\: J_f^2t^2}\dt
	    	\end{align}
	    	The quadratic coefficient $c_{\rm 1D}$ is the analog to the coefficient $c(\Delta)$ in Eq.~\ref{upEnvelope} and quantifies the damping of $\mathcal P_{k=0}(t)$. We show $c_{\rm 1D}$ as a function of temperature for several initial interaction strengths in Fig.~\ref{fig:smallNumericalDamping}.
	    	Many aspects of this figure are in good agreement with the analytical calculations presented in Sec.~\ref{momfin}. In particular, we see that for any fixed $U_i$ the damping is reduced for higher temperatures. 
			Furthermore the damping strength is independent of $U_i$ when the temperature $k_BT$ is much larger than $U_i$. For low temperatures, when the order parameter becomes substantial, there is a significant increase in $c_{\rm 1D}$. Finally $c_{\rm 1D}$ is larger for larger $U_i$, hence larger order parameter, for sufficiently high temperatures. 
			The most surprising difference to our analytical calculation is that we observe a finite amount of damping even for a non-interacting Fermi gas. We believe that this damping occurs, because we use a small system size and the canonical ensemble, where even a non-interacting Fermi gas is correlated. 

	    	In summary, our small numerical calculations show, in agreement with our analytical calculations, that the quadratic coefficient $c_{\rm 1D}$ approximately follows the value of the order parameter.

	\end{appendix}


	\bibliography{bib_paper}

\end{document}